\documentclass[twocolumn,journal]{IEEEtran}
\usepackage{graphicx}
\usepackage{amsmath}
\usepackage{amssymb}
\usepackage{color}
\usepackage{multicol}
\usepackage{amsthm}
\usepackage[usenames,dvipsnames]{xcolor}
\usepackage{balance}
\usepackage{cite}
\usepackage{float}

    \newcommand\blfootnote[1]{%
  \begingroup
  \renewcommand\thefootnote{}\footnote{#1}%
  \addtocounter{footnote}{-1}%
  \endgroup
}
\begin{document}
\title{Maximum Throughput of a Secondary User Cooperating with an Energy-Aware Primary User}
\author{ Ahmed El Shafie$^\dagger$, Ahmed Sultan$^\star$, Tamer Khattab$^*$\\
\small \begin{tabular}{c}
$^\dagger$Wireless Intelligent Networks Center (WINC), Nile University, Giza, Egypt. \\
$^\star$Department of Electrical Engineering, Alexandria University, Alexandria, Egypt.\\
$^*$Electrical Engineering, Qatar University, Doha, Qatar. \\
\end{tabular}
}

\date{}
\maketitle
\thispagestyle{empty}

\blfootnote{This paper was made possible by a NPRP grant 09-1168-2-455 from the
Qatar National Research Fund (a member of The Qatar Foundation). The
statements made herein are solely the responsibility of the authors.}
\pagestyle{empty}
\begin{abstract}
This paper proposes a cooperation protocol between a secondary user (SU) and a primary user (PU) which dedicates a free frequency subband for the SU if cooperation results in energy saving. Time is slotted and users are equipped with buffers. Under the proposed protocol, the PU releases portion of its bandwidth for secondary transmission. Moreover, it assigns a portion of the time slot duration for the SU to relay primary packets and achieve a higher successful packet reception probability at the primary receiver. We assume that the PU has three states: idle, forward, and retransmission states. At each of these states, the SU accesses the channel with adaptive transmission parameters. The PU cooperates with the SU if and only if the achievable average number of transmitted primary packets per joule is higher than the number of transmitted packets per joule when it operates alone. The numerical results show the beneficial gains of the proposed cooperative cognitive protocol.
\end{abstract}
\section{Introduction}
Cognitive usage of electromagnetic spectrum has been identified
as a useful means for efficient utilization of the scarce
radio resources. Under such paradigm, the
spectrum owned by the primary/licensed system is opportunistically
accessed by the secondary/unlicensed system using cognitive radio
technology.

Energy efficient communications has received significant attention recently \cite{li2011energy,phylayer,wang2006survey,crosslayer,imperfectcsi}. The problem of energy-efficient design was investigated from physical layer standpoint \cite{phylayer}, medium-access-control (MAC) layer standpoint \cite{wang2006survey}, and cross-layer standpoint \cite{crosslayer}. Cooperative communication is a method for achieving energy-efficient data transmission. Cooperative terminals act as spatially distributed antennas to provide alternative paths of the signal to the destinations. This may lead to significant reduction in transmit power and a better use of communication resources.

Combining cognitive radio technology and cooperative communications has been investigated in many papers, e.g. \cite{simeone,khattab,erph,su2011active}. In \cite{simeone} the secondary transmitter is used as a relay for the undelivered packets of the primary transmitter. The secondary user (SU) optimizes its power to expand the queueing stability region of the network.
In \cite{khattab}, the authors assumed that the cognitive transmitter is allowed to use the channel whenever the primary queue is empty. The SU cannot access the channel with its own data packets unless the relaying queue becomes empty. The secondary relays a certain fraction of the primary undelivered packets to minimize the secondary packets average delay subject to a power
budget for the relayed primary packets.
The maximum stable throughput of a secondary terminal sharing the channel with a primary terminal under multi-packet reception channel model was characterized in \cite{erph}. Based on the proposed scheme, the SU transmits its packets with some access probability based on the primary state. In particular, when the primary user (PU) is inactive, the SU immediately accesses the channel. When the PU is active, the SU probabilistically accesses the channel.
A new cooperation protocol was proposed in \cite{su2011active}. At each time slot, the PU releases portion of its bandwidth for the cognitive radio (CR) user and half of its time slot duration. During the first half of the time slot, the CR user receives the PU data. Afterwards, it amplifies-and-forwards the received packet during the second half of the time slot. The primary and secondary transmitters are assumed to be bufferless and to have a complete knowledge of the instantaneous transmit channel state information (CSI).

In this paper, we consider a cognitive network with an energy-aware primary terminal and a secondary terminal.
We propose the following cooperation protocol between the PU and the SU.
The PU releases portion of its bandwidth for secondary transmission. Moreover, it assigns a portion of the time slot duration for the SU to relay primary packets and achieve a higher successful packet reception probability at the primary receiver. The SU spends portion of its energy/power for aiding the PU to achieve certain quality of service requirements characterized by primary queue stability and higher number of transmitted primary packets per unit energy. The SU may leverage the primary feedback signal broadcasted in a time slot to ascertain the primary state in the following time slot. The secondary transmission time and bandwidth in a time slot change according to the state of the PU.
 We do not assume the availability of the transmit CSI at the transmitters.

The contributions of this paper can be summarized as follows.
\begin{itemize}
\item We propose a new cooperative cognitive relaying protocol for buffered primary and secondary terminals. The proposed protocol allows bandwidth and time sharing between the primary and the secondary users. Furthermore, the proposed protocol leverages the primary feedback channel.
    \item We consider an energy-aware PU and investigate the average number of transmitted primary packets per unit energy.
 \item In contrast to most of the existing works, we assume two quality of service requirements for the PU. Specifically, we put a constraint on the required number of transmitted primary packets per time slot, and another constraint on the primary queue stability. Violating any of these constraints obviates cooperation between the users.
 \item We derive closed-form expressions for the service rates of queues and the average number of transmitted primary packets per joule without and with cooperation.
 \item We characterize the maximum secondary throughput of the proposed protocol under the aforementioned constraints.
     \end{itemize}

The remainder of this paper is organized as follows. In the next section,
we describe the system model adopted in this paper. The description of the proposed cooperative protocol is found in Section \ref{protocol}. In Section \ref{sec3}, we discuss the stable-throughput region of the proposed protocol. Numerical results and conclusions are provided in Section \ref{sec4}.
\section{SYSTEM MODEL} \label{w100}
 We consider a cognitive network with one PU and one SU. This can be seen as a part of a larger network with multiple primary bands operating under frequency division multiple-access. Each band is composed of one PU and one SU. For simplicity of presentation, we provide the analysis of one of the available primary bands. The SU is assumed to be equipped with $\mathcal{M}$ antennas. The PU is assumed to be an energy-aware terminal, which operates in a time-slotted fashion with slot duration of $T$ seconds and a total bandwidth of $W$ Hz.

  \subsection{MAC Layer}\label{w111}
Each user is equipped with an infinite capacity buffer for storing its incoming traffic denoted by $Q_n$, where $n$ is `${\rm p}$' for primary and `${\rm s}$' for secondary. The SU has an additional finite capacity queue for storing the relaying packet, denoted by $Q_{\rm ps}$. Under the proposed protocol, the relaying queue maintains one packet at most, as explained later. We consider time-slotted transmissions where all packets have the same size of $b$ bits. The arrivals at $Q_{n}$, $n\in\{\rm p,s\}$, are assumed to be independent and identically distributed Bernoulli random variables from slot to slot with mean arrival rate $\lambda_{n}\!\in\![0,1]$ packets per time slot. The arrivals are also mutually independent from terminal to terminal. For similar network model and queue assumptions, the reader is referred to \cite{simeone,khattab}.

 The retransmission mechanism is based on the feedback acknowledgement/negative-acknowledgement (ACK/NACK) messages. More specifically, at the end of the time slot, each destination sends a feedback message to inform the respective transmitter about the decodability status of the transmitted packet. If a packet is received correctly at the respective destination, an ACK is fed back to the respective transmitter. On the other hand, if a packet is received erroneously at the respective receiver, a NACK message is fed back to the respective transmitter. We assume that
all nodes in the system can hear the feedback ACK/NACK messages. Therefore, the primary feedback messages are overheard by the SU. The overhead for transmitting the ACK and NACK messages is assumed to be very small
compared to packet sizes. Furthermore, the errors in packet acknowledgement feedback are negligible \cite{rong2012cooperative}, which
is reasonable for short length ACK/NACK packets as low rate
strong codes can be employed in the feedback channel.

The PU is assumed to have three states: idle, forward, and retransmission. The PU is said to be `idle', when its queue is empty. In the `forward' states, the PU sends the packet at the head of its queue, {\it for the first time}, while in the `retransmission' states, the PU transmits packets that have been erroneously decoded at the primary receiver. The primary feedback is either ``ACK", if the primary destination decodes the primary packet correctly; ``NACK", if the primary decoder fails in decoding the packet; or ``nothing", if there is no primary transmission.

  Without cooperation, the SU does not gain any access to the primary spectrum. This assumption is motivated by the fact that the primary system owns the spectrum. For the SU to gain access to the spectrum, it should provide an economic incentive or aid in enhancing the performance of the PU \cite{su2011active,chang2011cooperative,6504211}. In this work, we consider performance enhancement incentives.

  The PU cooperates with the SU if and only if cooperation yields more average number of transmitted packets per joule than when it operates alone. If cooperation is beneficial for the PU, it releases portion of its time slot duration and bandwidth for the SU. The portions of bandwidth and time slot released by the PU differ from state to state. In particular, if the primary queue is empty, the PU remains idle and the SU transmits its own packet over the whole channel bandwidth, $W$. Thus, the released bandwidth for the SU in idle states is $W$ Hz. If the primary queue is in a forward state, the released bandwidth and time slot duration are $W_{\rm s}\!=\!W\!-\!W_{\rm p}$ Hz and $T_{\rm s,F}\!=\!T\!-\!T_{\rm p,F}$ seconds, respectively. If the primary queue is in a retransmission state, the released bandwidth and time slot duration are $W_{\rm s}\!=\!W\!-\!W_{\rm p}$ Hz and $T_{\rm s,R}\!=\!T\!-\!T_{\rm p,R}$ seconds, respectively. In both cases, the SU uses the released bandwidth for its own data packets transmission. In addition, the SU assists the primary system in delivering its packet to the primary receiver using the assigned time and the remainder of the bandwidth, $W_{\rm p}$, which is used for both primary transmission and secondary transmission of the primary packet. The immediate benefits of the proposed cooperation protocol for the PU are transmission time reduction and transmission
energy savings per time slot.
 \subsection{Physical (PHY) Layer}
Wireless links exhibit fading and are corrupted by additive white Gaussian
noise (AWGN). The fading is assumed to be stationary, with
frequency non-selective Rayleigh block fading. This means
that the fading coefficient of a certain link remains constant during one slot. We do not assume the availability of the CSI at the transmitting terminals. The channel
is assumed to be known perfectly only at the receivers. The AWGN at each receiving node is assumed to have zero mean and power spectral density $\mathcal{N}_\circ$ Watts per unit frequency.

Denote the primary transmitter as `${\rm p}$', the primary destination as `${\rm pd}$', the $\nu${\it th} antenna of the secondary transmitter as `${\rm s}_{\nu}$', $\nu\in \{1,2,\dots,\mathcal{M}\}$, and the secondary destination as `${\rm sd}$'.
Let ${\rm H}$ denote the state of the primary terminal, where ${\rm H}$ is `${\rm \circ}$' for idle state, `${\rm F}$' for forward state, and finally, `${\rm R}$' for retransmission state.
The probability of channel outage of the link between node $j$ and node $\rho$ (link $j \rightarrow \rho$), for a given state ${\rm H}\in\{\circ,{\rm F,R}\}$, is given by $\mathbb{P}^{\rm H}_{j,\rho}$. This outage probability is a function of the number of bits in
a data packet, the slot duration, the transmission bandwidth, the
transmit powers, number of receive and transmit antennas, the average channel gains, and the state of the PU as detailed in
Appendix~A.

We assume one secondary antenna is used for transmission, and {\it all} antennas are used for channel sensing and data reception. Intuitively, increasing the expected link gain, $\sigma_{j,\rho}$ for the link between $j$ and $\rho$, decreases the outage probability of that link and, consequently, enhances queues' service rates. This is clear from the outage probability expression (\ref{ghj}) in Appendix~A. Based on this, the SU transmits to the primary or secondary receivers with the antennas having the highest expected link gain to the respective receivers. Let us assume, without loss of generality, that the link between the $i${\it th} secondary antenna and the secondary destination has the highest expected link gain among all the links between the SU's antennas and the secondary destination, and the link between the $m${\it th} secondary antenna and the primary destination has the highest expected link gain among all the links between the SU's antennas and the primary destination. The SU transmits a relaying packet using the $m${\it th} antenna, and transmits its own packet using the $i${\it th} antenna. Note that there possibly concurrent transmissions occur over orthogonal frequency subbands.

For proper operation of the proposed protocol, we assume that the minimum $W_{\rm p}T_{\rm p,F}$ is $\mathcal{E}$. This value is required for almost unity successful decoding of the primary packet at the SU terminal under the availability of $\mathcal{M}$ antennas and for certain channels parameters. In particular, $\mathcal{E}$ is designed such that the required outage probability of the link ${\rm p\rightarrow s}$ is at most $\mathcal{Q}$, where $\mathcal{Q}\ll 1$. Using the expression of probability of channel outage of link ${\rm p\rightarrow s}$ in Appendix A, we can find an expression for $\mathcal{E}$.
Assume that $P^{j}$ is the transmit power per unit frequency used by node $j$. Letting $\sigma_{{\rm p},{\rm s}_{\nu}}=\sigma_{{\rm p},{\rm s}}\ \forall {\nu}$ and noting that $\gamma_{{\rm p},{\rm s}_{\nu}}=P^{\rm p}/\mathcal{N}_\circ=\gamma_{{\rm p},{\rm s}}$ (equal for all receiving antennas), the outage probability (\ref{ccccvvv}) in Appendix A can be upper bounded as follows:
\begin{equation}
\mathbb{P}^{\rm F}_{{\rm p},{\rm s}}\!\le\!\Big[1-\exp(-\frac{2^{\frac{b}{W_{\rm p}T_{\rm p,F} }}\!-1}{\sigma_{{\rm p},{\rm s}}\gamma_{{\rm p},{\rm s}}})\Big]^\mathcal{M}\le \mathcal{Q}
\end{equation}
where the first inequality holds strictly to equality when the SU decodes separately the received copies of the primary packet at each antenna.
After some mathematical manipulations, we get
\begin{equation}
W_{\rm p}T_{\rm p,F}\!\ge\!{\frac{b}{\log_2\Big[1-{\sigma_{{\rm p},{\rm s}}\gamma_{{\rm p},{\rm s}}}\ln(1-\mathcal{Q}^{\frac{1}{\mathcal{M}}})\Big] }}\!=\!\mathcal{E}
\end{equation}
From the expression of $\mathcal{E}$, increasing the number of antennas reduces the required $\mathcal{E}$. Furthermore, as the received SNR, ${\sigma_{{\rm p},{\rm s}}\gamma_{{\rm p},{\rm s}}}$, increases, the signal quality increases, and $\mathcal{E}$ needed for a negligible decoding failure of the primary packet at the SU decreases as well. Finally, increasing the packet size, $b$, increases the required $\mathcal{E}$.
\section{Proposed Cooperative Cognitive Protocol}\label{protocol}
One of the most important functions for the SU is to ascertain the primary state at each time slot. The SU can discern the state of the PU via observing the primary feedback channel and/or channel sensing. In particular, if there is no primary feedback at the end of the previous time slot due to PU's inactivity, or if the SU hears an ACK, the SU senses the channel to discern the current state of activity of the PU. If the previous primary feedback is a NACK feedback, the SU does not sense the channel because it knows with certainty that the PU will be active in the current time slot. It is worth pointing out here that since the SU's operation is based on the channel sensing outcomes, the time assigned for channel sensing, $\tau$, is less than or equal to the primary transmission $T_{\rm p,F}$. That is, the domain of $T_{\rm p,F}$ is $[\tau,T]$.

Let us assume without loss of generality that the primary feedback received at the end of the previous time slot was not a NACK. In the current time slot, the SU must sense the channel for $\tau$ seconds to discern the state of activity of the PU. When the primary queue is empty, the SU transmits the packet at the head of its own traffic queue, $Q_{\rm s}$, with transmission rate $b/(T-\tau)$ and transmission bandwidth $W$ Hz. When the primary queue is nonempty, i.e, the PU is in the forward state, it releases $W_{\rm s}\!=\!W\!-\!W_{\rm p}$ for the SU to be used for its own data transmission. Hence, the SU's transmission rate is still $b/(T-\tau)$, but the transmission bandwidth becomes $W_{\rm s}\!\le\!W$ Hz. The PU and the SU use the remainder of the bandwidth, $W_{\rm p}\!=\!W\!-\!W_{\rm s}$, for the transmission of the primary packet over two non-overlapped time intervals. In particular, the PU sends its data packet over the time interval $[0,T_{\rm p,F}]$, at the same time, the SU attempts to decode the packet. If the SU is able to decode the primary packet, the packet is added to the relaying queue, $Q_{\rm ps}$. As mentioned earlier, we assume here that the SU always correctly decodes primary transmission by exploiting the spatial diversity provided by its multiple antennas. Over the remainder of the time slot, i.e., over the time interval $[T_{\rm p,F},T]$, the SU retransmits the primary packet to the primary receiver. At the end of the time slot, the primary destination decodes separately the received primary packet versions. If the primary destination decodes the primary packet successfully, the primary destination sends an ACK to inform the transmitters about the successful decoding of the primary packet. The primary packet is then dropped from both the primary and the relaying queues.

\begin{figure}[t]
\centering
  \includegraphics[width=1\columnwidth]{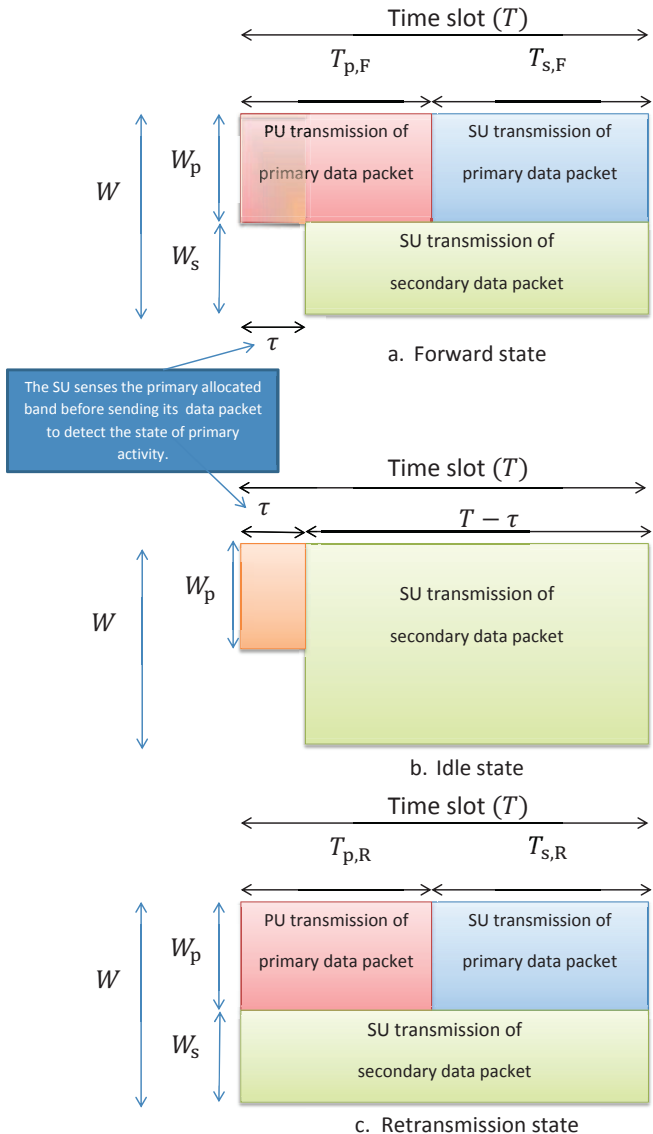}\\
  \caption{System operation and resource sharing at each of the primary states. In the figure, $T_{\rm s,F}=T-T_{\rm p,F}$ and $T_{\rm s,R}=T-T_{\rm p,R}$.}\label{syop}
\end{figure}

If the primary destination fails in decoding the primary packet at the first transmission, a NACK is sent by the primary destination to inform both the PU and the SU about the decoding failure of the primary packet. The retransmission of the primary packet starts at the following time slots until an ACK is issued by the primary destination. Since the SU does not sense the channel, and it knows with certainty that the PU is active, the SU transmits its own data packet over whole slot duration. Moreover, since the SU already has the primary packet in its buffer from the previous time slot, it does not have to receive/decode the primary packet again. During the retransmission states, the PU and the SU split the time slot into two portions as in the forward states such that the primary transmission takes place over $[0,T_{\rm p,R}]$, while the secondary transmission of the primary packet takes place over $[T_{\rm p,R},T]$. The system operation at each of the primary states is shown in Fig. \ref{syop}. Note that since there is no channel sensing in retransmission states, the primary transmission duration can take any value from $0$ to $T$. Hence, the domain of $T_{\rm p,R}$ is  $[0,T]$.

It is worth noting that the proposed
protocol requires time-synchronization of the cognitive radio system to the primary system. That can be ensured
via a beacon channel \cite{krikidis2009protocol}.

We assume that the misdetection and false-alarm probabilities are negligible (the same assumption is found in \cite{khattab,krikidis2009protocol}).
This means that the sensing duration $\tau$ is large enough to gather enough statistically independent samples using the SU's multiple antennas such that the error probabilities become almost equal to zero.

\section{Stability Analysis}\label{sec3}

 We adopt a late arrival model where new arrivals will not be served at the arriving slot even if the queue is empty \cite{sadek}.
 Denote by $A^t_{\zeta}$ the number of arrivals to queue $Q_{\zeta}$, ${\zeta}\in\{\rm p,s,ps\}$,
at time slot $t$, and $D^t_{\zeta}$ the number of departures from queue $Q_{\zeta}$ at time slot $t$. The queue length evolves according to the following form:
\begin{equation}
    Q_{\zeta}^{t+1}=(Q_{\zeta}^t-D^t_{\zeta})^+ +A^t_{\zeta}
\end{equation}
where $(z)^+$ denotes $\max(z,0)$.
A fundamental performance measure of a communication network is the stability of its queues. Stability can be defined as follows \cite{sadek}.

\emph{Definition:} Queue $Q_{\zeta}^t$, ${\zeta}\in\{\rm p,s,ps\}$, is stable, if
\begin{equation}\label{stabilityeqn}
    \lim_{y \rightarrow \infty} \lim_{t \rightarrow \infty  }{\rm Pr}\{Q_{\zeta}^t<y\}=1
\end{equation}
If the arrival and service processes are strictly stationary, then we can apply Loynes theorem to check for stability conditions \cite{loynes1962stability}. This theorem states that if the arrival process and the service process of a queue are strictly stationary processes, and the average service rate is greater than the average arrival rate of the queue, then the queue is stable. If the mean service rate is lower than the mean arrival rate, then the queue is unstable \cite{sadek}.
\subsection{Non-cooperative Users}

As mentioned earlier, without cooperation, the SU cannot access the channel and the PU transmits a lone at the beginning of the time slot and over the whole slot duration if its queue is nonempty. The primary queue mean service rate is equal to the complement of the channel outage between the PU and its respective receiver. Since the PU is an energy-aware terminal, at any given primary arrival rate, it maximizes its number of transmitted packets/joule. The PU transmits $b$ bits over $T$ seconds with transmission bandwidth $\mathbb{W}<W$, where $\mathbb{W}$ is assumed to be an optimizable parameter. The probability of successful primary transmission (i.e., the mean service rate) is given by
\begin{equation}
\mu_{\rm p,nc}=\overline{\mathbb{P}_{{\rm p},{\rm pd}}}=\exp\big(-\frac{2^{\frac{b}{{\mathbb{W}T}}}-1}{ \gamma_{{\rm p},{\rm pd}} \sigma_{{\rm p},{\rm pd}}}\big)
\label{cfffff}
\end{equation}
where $\overline{\mathcal{X}}\!=\!1\!-\!\mathcal{X}$. The probability of successful primary transmission is maximized when $\mathbb{W}$ is equal to $W$; hence, the maximum primary stable throughput is $\mu^{\max}_{\rm p,nc}~=~\exp~\big(~-~\frac{2^{\frac{b}{{WT}}}-1}{ \gamma_{{\rm p},{\rm pd}} \sigma_{{\rm p},{\rm pd}}}\big)$.

When the PU's queue is nonempty, the packet at the head of the queue is successfully received at the primary destination with probability $\mu_{\rm p,nc}$. Hence, the average number of transmitted packets per slot is $\mu_{\rm p,nc} {\rm Pr}\{Q_{\rm p}\ne 0\}$ packets/slot. Since the PU transmits with power $P^{\rm p}$ Watts/Hz over bandwidth $\mathbb{W}$ and with transmission time $T$ seconds, the average energy per slot is $P^{\rm p} \mathbb{W}T$ joule/slot.
Hence, the average number of transmitted primary packets per joule, $B_{\rm p,nc}$, is given by

\begin{equation}
\begin{split}
\label{muscccccx}
B_{\rm p,nc}&\!=\!\frac{1}{P^{\rm p}T \mathbb{W}}\mu_{\rm p,nc}{\rm Pr}\{Q_{\rm p}\ne 0\}\!
   \end{split}
\end{equation}
If the primary queue is stable, i.e., $\mu_{\rm p,nc}>\lambda_{\rm p}$, the probability of the primary queue being nonempty is given by ${\lambda_{\rm p}}/\mu_{\rm p,nc}$. Substituting with ${\rm Pr}\{Q_{\rm p}\ne 0\}\!=\!{\lambda_{\rm p}}/\mu_{\rm p,nc}$ into (\ref{muscccccx}), the number of transmitted primary packets per joule when the PU operates lonely is given by
\begin{equation}
\begin{split}
\label{musccccc}
B_{\rm p,nc}&\!=\!\frac{\lambda_{\rm p}}{P^{\rm p}T \mathbb{W}}
   \end{split}
\end{equation}
The maximum average number of transmitted primary packets per joule is obtained via solving the following optimization problem:
\begin{equation}
\begin{split}
\underset{\substack{\mathbb{W}}}{\max.} \,\,\ \!B_{\rm p,nc}=\frac{\lambda_{\rm p}}{P^{\rm p}T \mathbb{W}},
\,\,{\rm s.t.} \,\,\,\,\ \lambda_{\rm p} \le \mu_{\rm p,nc}, \ \mathbb{W} \in [0,W]
\end{split}
\label{opixxxx}
\end{equation}
The optimization problem can be converted to the following linear program:
\begin{equation}
\begin{split}
\underset{\substack{\mathbb{W}}\in[0,W]}{\min.} \  \mathbb{W},
\,\,{\rm s.t.} \,\  {{\mathbb{W}}}\!\ge\! \frac{b} {T\log_2\Big(1\!-\!{ \gamma_{{\rm p},{\rm pd}} \sigma_{{\rm p},{\rm pd}}}\ln(\lambda_{\rm p})\Big)}\!=\!{{\mathbb{W}}}_{\min}
\end{split}
\label{opixxxx2}
\end{equation}
It is straightforward to show that the optimal primary transmission bandwidth and the maximum average number of transmitted primary packets per joule are given by
\begin{equation}
\begin{split}
  {{\mathbb{W}}^*}\!=\!{{\mathbb{W}}}_{\min}, \ \!B^{\max}_{\rm p,nc}=\frac{\lambda_{\rm p}}{P^{\rm p}T {\mathbb{W}}^*}
\end{split}
\end{equation}
with $\lambda_{\rm p}\!\le\!\mu^{\max}_{\rm p,nc}$. From the optimal solution, we note that ${{\mathbb{W}}}_{\min}$ increases with $\lambda_{\rm p}$. This is expected as the primary queue starts to saturate and the needed bandwidth expands to increase the successful probability of the primary transmissions, which in turn maintain the stability of the primary queue. Furthermore, for a given $\lambda_{\rm p}$, the bandwidth used for transmission decreases with $\gamma_{{\rm p},{\rm pd}} \sigma_{{\rm p},{\rm pd}}$ and $T/b$. This is because $\mu_{\rm p}$ must be at least equal to $\lambda_{\rm p}$ for maintaining the primary queue stability.
\begin{figure}[t]
\centering
  \includegraphics[width=1\columnwidth]{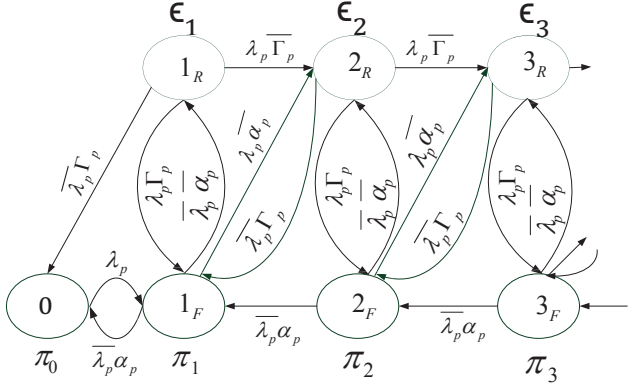}\\
  \caption{Markov chain of the PU for the proposed protocol. State self-transitions are not
depicted for visual clarity. Note that $\overline{\mathcal{X}}\!=\!1\!-\!\mathcal{X}$.}\label{figxx}
\end{figure}
\subsection{Cooperative Users}
\subsubsection{Markov Chain Analysis of The PU's Queue}
The Markov chain representing the PU's queue when the PU cooperates with the SU is shown in Fig. \ref{figxx}. The probability of the primary queue
having $k$ packets and transmitting for the first time is $\pi_k$, where
$F$ in Fig. \ref{figxx} denotes first transmission. The probability of the
primary queue having $k$ packets and retransmitting is $\epsilon_k$, where $R$ in
Fig. \ref{figxx} denotes retransmission.
 Let $\alpha_{\rm p}$ denote the probability of primary successful transmission in case of first transmission, and $\Gamma_{\rm p}$ denote the probability of successful transmission in case of retransmission. Recall that the $m${\it th} antenna is the antenna element within secondary array that has the maximum average link gain to the primary destination. Since the primary receiver decodes separately the received copies of the primary packet from each user, the probability of primary packet correct reception in forward or retransmission states is equal to one minus the probability of both ${\rm p\rightarrow pd}$ and ${{\rm s}_m {\rm \rightarrow \rm pd}}$ being in outage simultaneously. Hence, $\alpha_{\rm p}$ and $\Gamma_{\rm p}$ are given by
\begin{equation}
\begin{split}
    \alpha_{\rm p}&\!=\!1-\mathbb{P}^{\rm F}_{\rm p,pd} \mathbb{P}^{\rm F}_{{\rm s}_m,{\rm pd}}, \ \Gamma_{\rm p}\!=\!1-\mathbb{P}^{\rm R}_{\rm p,pd} \mathbb{P}^{\rm R}_{{\rm s}_m,{\rm pd}}
    \end{split}
\end{equation}
where $\mathbb{P}^{\rm F}_{\rm p,pd}$, $\mathbb{P}^{\rm F}_{{\rm s}_m,{\rm pd}}$, $\mathbb{P}^{\rm R}_{\rm p,pd}$, and $\mathbb{P}^{\rm R}_{{\rm s}_m,{\rm pd}}$ follow (\ref{ghj}) with ${\rm H\in\{F,R\}}$, $r^{\rm H}_{{\rm s}_m}=b/T_{\rm s,H}$, $r^{\rm H}_{{\rm p}}=b/T_{\rm p,H}$ and $W_{{\rm s}_m, \rm pd}\!=\!W_{\rm p}$.

 Solving the state balance equations, we can obtain the state probabilities which are provided in Table \ref{table}. The detailed solution is omitted here due to space limitations. The stability of the PU's queue is attained when the
probability of its queue being empty is greater than zero. This can be guaranteed when $\lambda_{\rm p}<\eta$, where $\eta$ is defined in Table \ref{table}.

\begin{table}[t]
\begin{center}
\caption{State probabilities for the PU's Markov chain.}
\label{table}
\begin{tabular}{ |@{}c@{}|@{}c@{}|@{}c@{}|@{}c@{}|  }
    \hline
           \hbox{ Parameter} & \hbox{Value}& \hbox{ Parameter} & \hbox{Value}\\[6pt]\hline
    $\eta$ & $\lambda_{\rm p}\alpha_{\rm p}+\overline{\lambda_{\rm p}}\Gamma_{\rm p}$&$\pi_{\circ}$& $\frac{\eta-\lambda_{\rm p}}{\Gamma_{\rm p}}$  \\[10pt]\hline
   &  & $\psi$ & $\overline{\lambda_{\rm p}} \eta$ \\[5pt]\hline $\pi_1$ & $\pi_{\circ}\tfrac{\lambda_{\rm p}}{\psi}(\lambda_{\rm p}+\overline{\lambda_{\rm p}}\Gamma_{\rm p})$ &
  $\epsilon_1$ & $\pi_{\circ}\frac{\lambda_{\rm p}\overline{\alpha_{\rm p}}}{\eta}$ \\[6pt]\hline $\pi_k,k\geq 2$ & $\pi_{\circ}\frac{\lambda_{\rm p}\overline{\alpha_{\rm p}}}{\overline{\eta}^2} \bigg[ \frac{\lambda_{\rm p} \overline{\eta}}{\psi}  \bigg]^k$&
  $\epsilon_k,k\geq 2$ & $\pi_{\circ}\frac{\overline{\lambda_{\rm p}}\overline{\alpha_{\rm p}}}{\overline{\eta}^2} \bigg[ \frac{\lambda_{\rm p} \overline{\eta}}{\psi}  \bigg]^k$ \\ [8pt]\hline $\sum_{k=1}^{\infty}\pi_k$ & $\pi_{\circ}\frac{\lambda_{\rm p}\Gamma_{\rm p}}{\eta-\lambda_{\rm p}}=\lambda_{\rm p}$&
   $\sum_{k=1}^{\infty}\epsilon_k$ & $\pi_{\circ}\frac{\lambda_{\rm p}}{\eta-\lambda_{\rm p}}\overline{\alpha_{\rm p}}=\frac{\lambda_{\rm p} }{\Gamma_{\rm p}}\overline{\alpha_{\rm p}}$ \\[6pt]\hline
\end{tabular}
\end{center}
\end{table}
\bigskip
\subsubsection{Secondary Throughput and Average Number of transmitted Primary Packets per joule}
The packet at the head of $Q_{\rm s}$ is served if, for a given primary state, the channel between the $i${\it th} antenna\footnote{Recall that the $i${\it th} antenna is the antenna element within secondary array that has the maximum average link gain to the secondary destination.} and the secondary receiver is not in outage. The average service rate of the secondary queue, denoted by $\mu_{\rm s}$, is then given by
\begin{equation}
\begin{split}
\label{mus}
  \mu_{{\rm s}}&\!=\!\pi_o \overline{\mathbb{P}^{ \circ}_{{\rm s}_i,{\rm sd}}} \!+\!(\sum_{k\ge1} \pi_k) \overline{\mathbb{P}^{\rm F}_{{\rm s}_i,{\rm sd}}} \!+\!(\sum_{k\ge1} \epsilon_k) \overline{\mathbb{P}^{\rm R}_{{\rm s}_i,{\rm sd}}}\\\!&=\!\pi_o \overline{\mathbb{P}^{ \circ}_{{\rm s}_i,{\rm sd}}} +\lambda_{\rm p} \overline{\mathbb{P}^{\rm F}_{{\rm s}_i,{\rm sd}}} +\frac{\lambda_{\rm p} \overline{\alpha_{\rm p}} }{\Gamma_{\rm p}}\overline{\mathbb{P}^{\rm R}_{{\rm s}_i,{\rm sd}}}
   \end{split}
\end{equation}
where the probabilities $\pi_\circ$, $\sum_{k\ge1} \pi_k$ and $\sum_{k\ge1} \epsilon_k$ are given in Table \ref{table}.

Under the proposed protocol, the average number of transmitted primary packets per joule, $B_{\rm p,c}$, is given by
\begin{equation}
\label{muccc}
B_{\rm p,c} \!=\! \left\{ \begin{array}{ll}
         \!\frac{\alpha_{\rm p}(\sum_{k\ge1} \pi_k)}{P^{\rm p}W_{\rm p}T_{\rm p,F}}  \!+\!\frac{\Gamma_{\rm p} (\sum_{k\ge1} \epsilon_k)} {P^{\rm p}W_{\rm p}T_{\rm p,R}}\!=\!\frac{ \frac{ \alpha_{\rm p}}{T_{\rm p,F}}\!+\!\frac{  \overline{\alpha_{\rm p}}}{T_{\rm p,R}} }{W_{\rm p}P^{\rm p}/\lambda_{\rm p}} & \mbox{if $T_{\rm p,R}\!>\!0$};\\
        \frac{\alpha_{\rm p} (\sum_{k\ge1} \pi_k)}{P^{\rm p}W_{\rm p} T_{\rm p,F}}\!=\!\frac{\alpha_{\rm p} \lambda_{\rm p}}{P^{\rm p}W_{\rm p} T_{\rm p,F}} & \mbox{if $T_{\rm p,R}\!=\!0$}.
        \end{array} \right.\\
        \end{equation}
The expression in (\ref{muccc}) is explained as follows. First, we note that $T_{\rm p,F}$ is at least equal to $\min\{\tau,\mathcal{E}/W_{\rm p}\}\!>\!0$, whereas $T_{\rm p,R}$ can be assigned any value between $0$ and $T$. If $T_{\rm p,R}\!=\!0$, this means that the PU does not spend any energy for packets retransmissions, and the responsibility of delivering the primary packets at retransmission states is entirely given to the SU. Consequently, the value of $T_{\rm p,R}$ partitions $B_{\rm p,c}$ into two sets. Given that the probability of the PU being in a forward state is $\sum_{k\ge1} \pi_k$, the packet at the head of the primary queue is received successfully at the primary destination with probability $\alpha_{\rm p}$. Hence, the number of successfully transmitted primary packets per slot is $(\sum_{k\ge1} \pi_k)\alpha_{\rm p}$ packets/slot. Since the PU transmits with power $P^{\rm p}$ Watts/Hz over bandwidth $W_{\rm p}$ Hz and with transmission time $T_{\rm p,F}$ seconds, the average primary transmit energy per slot is $P^{\rm p}W_{\rm p}T_{\rm p,F}$ joule/slot.

In a similar fashion, if the PU is in a retransmission state, given that the probability of the PU being in a retransmission state is $\sum_{k\ge1} \epsilon_k$, the average number of successfully transmitted primary packets per slot is $(\sum_{k\ge1} \epsilon_k)\Gamma_{\rm p}$ packets/slot, and the average primary transmit energy per slot is $P^{\rm p}W_{\rm p}T_{\rm p,R}$ joule/slot. If $T_{\rm p,R}\!=\!0$, the PU does not spend any energy for its packet delivery during retransmission states and the average primary transmit energy per slot is zero; hence, $B_{\rm p,c}$ is given by the value explained in the forward states. Specifically, $B_{\rm p,c}\!=\!\frac{\alpha_{\rm p} (\sum_{k\ge1} \pi_k)}{W_{\rm p}P^{\rm p} T_{\rm p,F}}$ packets/joule for $T_{\rm p,R}=0$.
\subsubsection{Problem Formulation}
For cooperation to be established, $B_{\rm p,c}$ must be strictly greater than $B^{\max}_{\rm p,nc}$, i.e., $B_{\rm p,c}> B^{\max}_{\rm p,nc}$. We characterize the maximum secondary throughput under such quality of service condition. The maximum secondary throughput of the proposed protocol is characterized by the closure of the rate pairs $(\lambda_{\rm p},\mu_{\rm s})$. To obtain this closure, we maximize the mean service rate of the SU under constraints on the stability of system's queues\footnote{The stability of the relaying queue is obvious as it contains at most one packet.}, $B_{\rm p,c}> B_{\rm p,nc}^{\max}$, and $W_{\rm p}T_{\rm p,F}\ge \mathcal{E}$. The optimization problem is stated as:
\begin{equation}
\begin{split}
\underset{\substack{W_{\rm p}, T_{\rm p,F}, T_{\rm p,R}}}{\max.} \,\,\ \!\mu_{{\rm s}},\\
\,\,{\rm s.t.} \,\,\,\,\ \lambda_{\rm p} &\le \eta, \ B_{\rm p,c}> B_{\rm p,nc}^{\max}, \  W_{\rm p}T_{\rm p,F}\ge \mathcal{E} \\ W_{\rm p}& \in [0,W], \ T_{\rm p,F} \in [\tau, T], \ T_{\rm p,R} \in [0, T]
\end{split}
\label{opi}
\end{equation}
The optimization problem (\ref{opi}) is solved numerically.\footnote{The optimal parameters obtained via solving the optimization problem (\ref{opi}) are announced to both users so that $W_{\rm p}$, $T_{\rm p,F}$ and $T_{\rm p,R}$ are known at the PU and the SU before actual operation.} In particular, we make a grid search over the optimization parameters. It should be pointed out here that the impact of the number of antennas, $\mathcal{M}$, affects the feasible set of (\ref{opi}) as it changes the value of $\mathcal{E}$. As is shown in Appendix B, increasing $\mathcal{M}$ decreases $\mathcal{E}$ which in turn may reduce the required value of bandwidth assigned for primary packet transmission. Consequently, the bandwidth assigned for the SU when the PU is active may increase, which boosts the secondary throughput.

If the distance between the PU and its respective receiver is long or the direct link is in deep shadowing due to surrounding physical obstacles, the primary direct link will be disconnected \cite{riihonen2011mitigation,bletsas2006simple}. This means that $\sigma_{\rm p,pd}\approx0$. In this case, the time assigned to the PU for transmission must be minimized. This is because there is no beneficial gain for the PU to use most of the time slot for delivering the packet to its receiver. Instead assigning most of the time slot for the SU to relay that packet may be more beneficial if it has relatively better channel quality to the primary destination. Thus, the optimal parameters are given by
\begin{equation}
\begin{split}
T_{\rm p,F}=\max\{\frac{\mathcal{E}}{W_{\rm p} },\tau\},\  T_{\rm p,R} =0
\end{split}
\end{equation}
The problem reduces to an optimization problem with single optimization parameter, $W_{\rm p}$, and with one constraint, $\lambda_{\rm p}\le \eta$.
In this case, the successful probabilities of the primary packets at forward and retransmission states are
\begin{equation}
\begin{split}
\alpha_{\rm p}\!=\!\exp\bigg(\!-\!\frac{2^\frac{b}{W_{\rm p}(T\!-\!\max\{\frac{\mathcal{E}}{W_{\rm p} },\tau\} \!)}\!-\!1}{\gamma_{\rm s,pd}\sigma_{\rm s,pd}}\!\bigg), \ \Gamma_{\rm p}\!=\!\exp\bigg(\!-\!\frac{2^\frac{b}{W_{\rm p}T}\!-\!1}{\gamma_{\rm s,pd}\sigma_{\rm s,pd}}\!\bigg)
\end{split}
\end{equation}

The feasible set of the optimization problem is given by
\begin{equation}
\begin{split}
\lambda_{\rm p}\le \eta \Longleftrightarrow \frac{\lambda_{\rm p} }{\overline{\lambda_{\rm p}}}\le \frac{\Gamma_{\rm p}}{1-\alpha_{\rm p}}
\end{split}
\end{equation}
The optimization problem becomes:
\begin{equation}
\begin{split}
&\underset{\substack{W_{\rm p}\in[0,W]}}{\max.} \,\,\ \!\mu_{{\rm s}},\\ &
\,\,{\rm s.t.} \,\,\,\,\ \frac{\lambda_{\rm p} }{\overline{\lambda_{\rm p}}}\le \frac{\Gamma_{\rm p}}{1-\alpha_{\rm p}}
\end{split}
\label{opi2}
\end{equation}
This optimization problem is solved via a simple grid search over $W_{\rm p}$.
\section{Numerical Results and Conclusions}\label{sec4}
In this section, we provide numerical results to illustrate the gains of the proposed protocol. For sake of simplicity, we denote $\sigma_{{\rm s}_i,{{\rm sd}}}=\sigma_{{\rm s}_i,{{\rm sd}}}$ and $\sigma_{{\rm s}_m,{{\rm sd}}}=\sigma_{{\rm s},{{\rm pd}}}$.
The common parameters used to generate the figures are: $\mathcal{Q}= 10^{-8}$, $b=2000$ bits, $W=10$ MHz, $T=4\times 10^{-4}$ seconds, $\mathcal{N}_\circ=10^{-11}$ Watts/Hz, $P^{\rm p}= 10^{-10}$ Watts/Hz, $\sigma_{{\rm s},{{\rm sd}}}=0.1$, $\sigma_{{\rm p},{\rm pd}}=0.2$, $\sigma_{{\rm s},{{\rm pd}}}=0.5$, $\sigma_{{\rm p},{\rm s}_{\nu}}\!=\!\sigma_{{\rm p},{\rm s}}\!=\!1$ for all $\nu$, and $\tau=0.2T$.

\begin{figure}
  \centering
  \includegraphics[width=1\columnwidth]{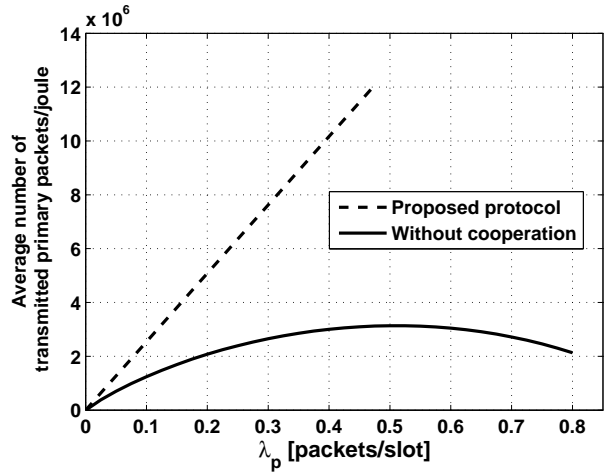}\\
   \caption{The primary average packets per joule versus $\lambda_{\rm p}$.}\label{fig7}
\end{figure}

In Fig. \ref{fig7}, we present the average number of transmitted primary packets per joule versus $\lambda_{\rm p}$ for the proposed protocol. The non-cooperation case is also plotted for comparison. The figure reveals that the average number of transmitted primary packets per joule of the proposed protocol is higher than the maximum average number of transmitted primary packets per joule of the non-cooperative case over most of the $\lambda_{\rm p}$ range. More specifically, over $\lambda_{\rm p}<0.475$ packets/slot, the proposed protocol outperforms the non-cooperative case. For $\lambda_{\rm p}\!\ge\!0.475$ packets/slot, $B_{\rm p,c}\le B^{\max}_{\rm p,nc}$; hence, the cooperation becomes non-beneficial for the PU and it ceases to use the SU for relaying. In this case, the optimization problem of the proposed protocol becomes infeasible. Hence, the SU does not gain any access to the channel. The parameters used to generate this figure are the common parameters, $\mathcal{M}=6$, and $P^{\rm s}=5\times 10^{-11}$ Watts/Hz.

\begin{figure}
  \centering
  \includegraphics[width=1\columnwidth]{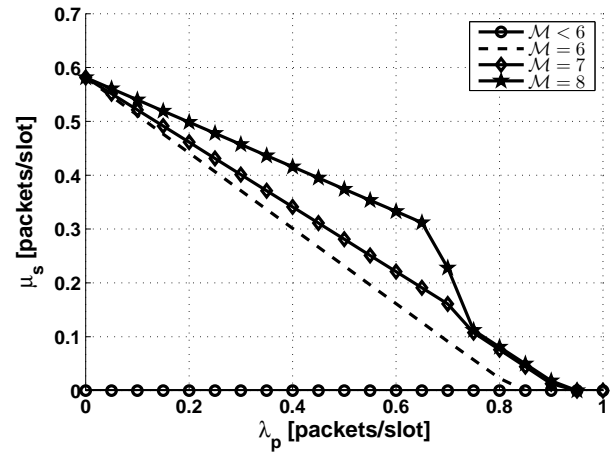}\\
   \caption{The impact of the number of antennas on the secondary throughput.}\label{fig6}
\end{figure}

\begin{figure}
  \centering
  \includegraphics[width=1\columnwidth]{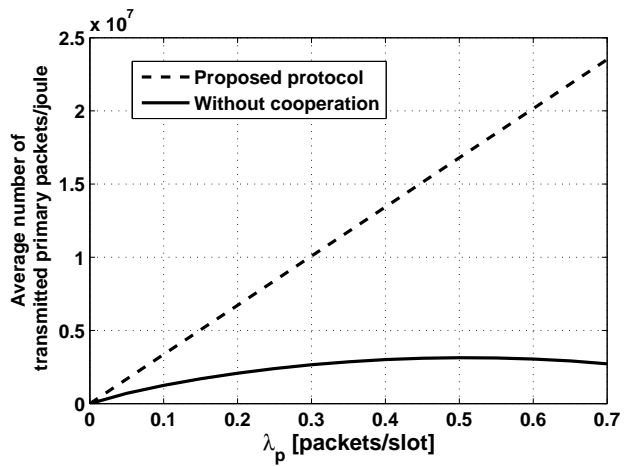}\\
   \caption{The number of transmitted primary packets per joule.}\label{fig1}
\end{figure}

Fig. \ref{fig6} demonstrates the impact of the number of secondary antennas on
the maximum secondary stable throughput. Under the used parameters, for $\mathcal{M}<6$, the SU cannot access the channel because the condition $W_{\rm p}T_{\rm p,F}\ge \mathcal{E}$, which guarantees unity probability of primary packet decoding at the SU, is not satisfied for the feasible range of $W_{\rm p}$ and $T_{\rm p,F}$. The figure also demonstrates the gains of increasing the number of secondary antennas which boosts the secondary throughput. The secondary throughput increases with $\mathcal{M}$ because as the number of antennas increases, the assigned bandwidth for the PU may be decreased due to the fact that $\mathcal{E}\!\le\! W_{\rm p}T_{\rm p,F}$ is monotonically decreasing with $\mathcal{M}$. Hence, the assigned bandwidth for the SU when the PU is active, $W\!-\!W_{\rm p}$, may increase, which in turn increases the probability of secondary packet successful decoding and the secondary throughput. The figure is generated using the common parameters, $\mathcal{M}=7$ and $P^{\rm s}=10^{-10}$ Watts/Hz. Fig. \ref{fig1} shows the number of primary transmitted packets per joule with and without cooperation. The figure is generated with the same parameters used to generate Fig. \ref{fig6} and with $\mathcal{M}=7$ antennas.
 From the figure, the gain of the proposed protocol over the noncooperation case is obvious. When $\lambda_{\rm p}=0.7$ packets/slot, the gain of the proposed protocol over the noncooperation case is almost 765\%.

\begin{figure}
  \centering
  \includegraphics[width=1\columnwidth]{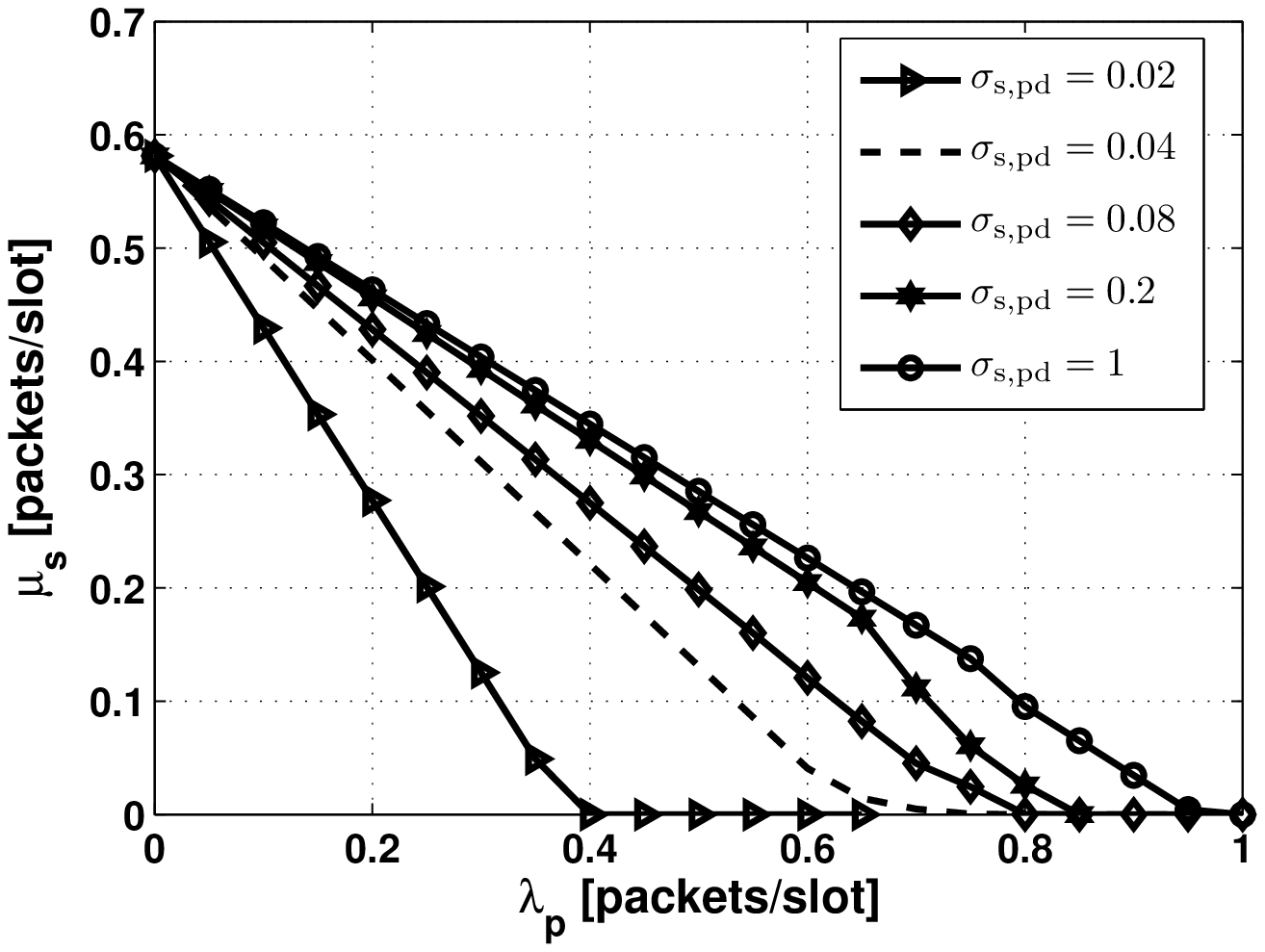}\\
   \caption{Maximum mean secondary service rate for the proposed system when the expected channel gain of the highest channel gain of the link connecting the $m${\it th} antenna and the primary destination varies.}\label{fig2}
\end{figure}

\begin{figure}
  \centering
  \includegraphics[width=1\columnwidth]{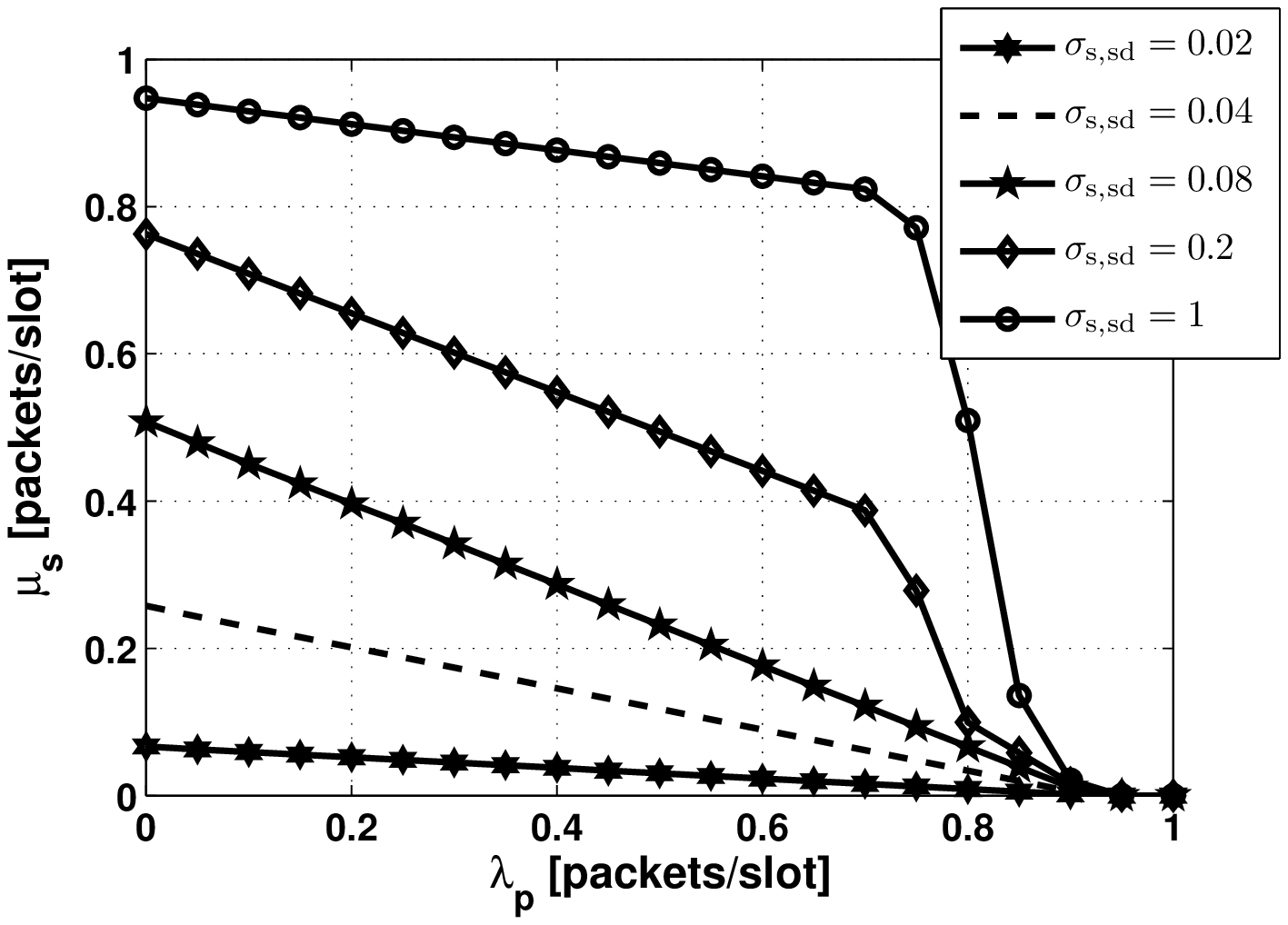}\\
   \caption{Maximum mean secondary service rate for the proposed system when the expected channel gain of the highest channel gain of the link connecting the $i${\it th} antenna and the secondary destination varies.}\label{fig3}
\end{figure}

\begin{figure}
  \centering
  \includegraphics[width=1\columnwidth]{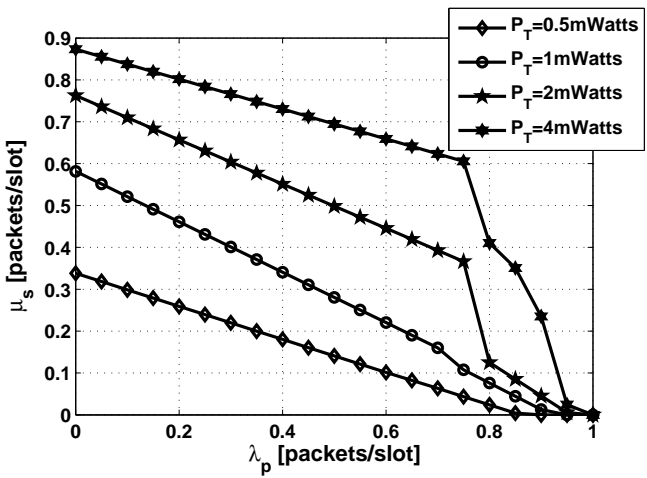}\\
   \caption{Maximum secondary service rate of the proposed protocol when secondary transmit power, $P_{\rm T}=P^{\rm s} W$, varies.}\label{fig4}
\end{figure}

  Fig. \ref{fig2} shows the maximum secondary throughput of the proposed system when the highest mean among the channels between the SU's antennas and the primary destination, $\sigma_{\rm s_m, pd}$, varies. It can be noticed that the feasible range of the primary arrivals increases with $\sigma_{\rm s_m, pd}$. Fig. \ref{fig3} presents the maximum secondary throughput of the proposed system when the highest mean among the channels between the SU's antennas and its respective destination, $\sigma_{\rm s_i, sd}$, varies. It can be noticed that the feasible range of the secondary arrivals increases with $\sigma_{\rm s_i, pd}$. From Figs. \ref{fig2} and \ref{fig3} we can conclude that for a given $\lambda_{\rm p}$, increasing $\sigma_{\rm s_i, sd}$ or $\sigma_{\rm s_m, pd}$ boosts the secondary throughput. Ditto for a given $\lambda_{\rm s}$, increasing $\sigma_{\rm s_i, sd}$ or $\sigma_{\rm s_m, pd}$ boosts the primary throughput. The parameters used to generate each of the figures are the common parameters, $\mathcal{M}=7$, $P^{\rm s}=10^{-10}$ Watts/Hz and the parameters in the legend of each figure. Fig. \ref{fig4} shows the maximum secondary stable throughput of the proposed system for different $P_T=WP^{\rm s}$. Increasing $P^{\rm s}$ boosts the maximum secondary stable throughput as well as the feasible primary mean arrival rate. This is because increasing the transmit power decreases the outage probability of the link. This can be noticed from the outage probability formula. The figure is generated using the common parameters and the parameters in the figure's legend.

%
%

\section*{Appendix A}
\subsection{Channels Outages}
 Let $g^t_{j,\rho}$ denote the channel gain between transmitting node $j$ and receiving node $\rho$ at instant $t$, where $j \in \{{\rm s}_{1},{\rm s}_{2},\dots,{\rm s}_{\mathcal{M}},{\rm p}\}$, $\rho \in \{{\rm s}_{1},{\rm s}_{2},\dots,{\rm s}_{\mathcal{M}},{\rm sd},{\rm pd}\}$ and $j\!\ne\! \rho$, which is exponentially distributed in case of Rayleigh fading channel with mean $\sigma_{j,\rho}$. Hereinafter, the time notation is omitted from all symbols for simplicity. Channel gains are independent from slot to slot and link to link.
\noindent Outage occurs when the transmission rate exceeds the channel capacity. Since the transmission rates of all nodes change from state to state, the outage probability should be parameterized by the state of the PU. The probability of channel outage of the link between node $j$ and node $\rho$, for a given state ${\rm H}\in\{\circ,{\rm F,R}\}$, is given by \cite{rong2012cooperative}
\begin{equation}
\begin{split}
\mathbb{P}^{\rm H}_{j,\rho}\!=\!{\rm Pr}\biggr\{\!r^{\rm H}_j \!>\! W^{\rm H}_{j,\rho} \log_{2}\left(1\!+\!\beta_{j,\rho}\right)\!\biggr\}\!=\! 1\!-\!\exp\bigg(-\!\frac{2^{\frac{r^{\rm H}_{j}}{W_{j,\rho}}}\!-\!1}{\gamma_{j,\rho}\sigma_{j,\rho}}\bigg)
\label{ghj}
\end{split}
\end{equation}
\noindent where $r^{\rm H}_j$ is the transmission rate of transmitter $j$ at state ${\rm H}$, $W^{\rm H}_{j,\rho}$ is the transmission bandwidth used for the communication between node $j$ and node $\rho$ at primary state ${\rm H}$, $\beta_{j,\rho}=\frac{P^{j} g_{j,\rho}}{\mathcal{N}_\circ}$ is the instantaneous signal-to-noise ratio (SNR), $\gamma_{j,\rho}=P^{j}/\mathcal{N}_\circ$ is the received SNR at node $\rho$ when the channel gain is equal to unity, and $P^{j}$ is the transmit power per unit frequency used by node $j$. Note that $W^{\rm H}_{\rm p,pd}\!=\!W^{\rm H}_{\rm s_{\nu},pd}\!=\!W_{\rm p}$, where ${\rm H\in\{F,R\}}$, $W^{\rm F}_{\rm s_{\nu},sd}\!=\!W^{\rm R}_{\rm s_{\nu},sd}\!=\!W_{\rm s}$ and $W^{\rm \circ}_{\rm s_{\nu},sd}\!=\!W$. It should be mentioned that the transmit power by any of the SU's antennas is $P^{\rm s}$. Hence, the received SNR at destination $\rho$, when the channel gain is unity, is $\gamma_{{\rm s}_{\nu},\rho}=\gamma_{{\rm s},\rho}=P^{\rm s}/\mathcal{N}_\circ$. We note that the outage probability (\ref{ghj}) decreases with $\sigma_{j,\rho}$.
\subsection{Secondary Transmission Rate}
The transmission rate of the secondary transmitter differs from state to state. If the PU is idle or sending a packet for the first time (forward state), the secondary rate is given by
\begin{equation}
r^{\rm F}_{{\rm s}_i}=r^{\circ}_{{\rm s}_i}=\frac{b}{T-\tau}
\label{r_i}
\end{equation}
In case of retransmission of a primary packet, the SU does not sense the channel because it knows with certainty that the PU is active. Hence, the secondary transmission rate is given by
\begin{equation}
r^{\rm R}_{{\rm s}_i}=\frac{b}{T}
\label{r_}
\end{equation}

\subsection{Primary Packet Decoding at the SU}
In forward states, the PU sends a packet of size $b$ bits over $[0,T_{\rm p,F}]$; hence, the primary rate is $r^{\rm F}_{\rm p}\!=\!b/T_{\rm p,F}$. We assume that the SU combines what it gets from each antenna\footnote{Recall that the receiving nodes have the CSI.}; hence, the probability of primary packet decoding failure by the SU is given by
\begin{equation}
\mathbb{P}^{\rm F}_{{\rm p},{\rm s}}={\rm Pr}\Big\{\!\sum_{\nu=1}^{\mathcal{M}} g_{{\rm p},{\rm s}_{\nu}}\gamma_{{\rm p},{\rm s}_{\nu}}<2^{\frac{b}{T_{\rm p,F} W_{\rm p} }}\!-1\!\Big\}\!
\label{ccccvvv}
\end{equation}
where $\gamma_{{\rm p},{\rm s}_{\nu}}=P^{\rm p}/\mathcal{N}_\circ$ and $P^{\rm p}$ is the primary transmit power. The closed-form of the probability in (\ref{ccccvvv}) can be found in \cite{scheuer1988reliability}.\footnote{This probability, $\mathbb{P}^{\rm F}_{{\rm p},{\rm s}}$, can be upper bounded by the probability of having $\mathcal{M}$ antennas each of which is independently decodes the primary packet.}
\bibliographystyle{IEEEtran}
   \balance
\bibliography{IEEEabrv,bibfile}

\begin{thebibliography}{10}
\providecommand{\url}[1]{#1}
\csname url@samestyle\endcsname
\providecommand{\newblock}{\relax}
\providecommand{\bibinfo}[2]{#2}
\providecommand{\BIBentrySTDinterwordspacing}{\spaceskip=0pt\relax}
\providecommand{\BIBentryALTinterwordstretchfactor}{4}
\providecommand{\BIBentryALTinterwordspacing}{\spaceskip=\fontdimen2\font plus
\BIBentryALTinterwordstretchfactor\fontdimen3\font minus
  \fontdimen4\font\relax}
\providecommand{\BIBforeignlanguage}[2]{{%
\expandafter\ifx\csname l@#1\endcsname\relax
\typeout{** WARNING: IEEEtran.bst: No hyphenation pattern has been}%
\typeout{** loaded for the language `#1'. Using the pattern for}%
\typeout{** the default language instead.}%
\else
\language=\csname l@#1\endcsname
\fi
#2}}
\providecommand{\BIBdecl}{\relax}
\BIBdecl

\bibitem{li2011energy}
G.~Y. Li, Z.~Xu, C.~Xiong, C.~Yang, S.~Zhang, Y.~Chen, and S.~Xu,
  ``Energy-efficient wireless communications: tutorial, survey, and open
  issues,'' \emph{IEEE Wireless Communications}, vol.~18, no.~6, pp. 28--35,
  2011.

\bibitem{phylayer}
E.~V. Belmega, S.~Lasaulce, and M.~Debbah, ``A survey on energy-efficient
  communications,'' in \emph{IEEE 21st International Symposium on PIMRC
  Workshops}, 2010, pp. 289--294.

\bibitem{wang2006survey}
L.~Wang and Y.~Xiao, ``A survey of energy-efficient scheduling mechanisms in
  sensor networks,'' \emph{Mobile Networks and Applications}, vol.~11, no.~5,
  pp. 723--740, 2006.

\bibitem{crosslayer}
G.~Miao, N.~Himayat, Y.~G. Li, and A.~Swami, ``Cross-layer optimization for
  energy-efficient wireless communications: a survey,'' \emph{Wireless
  Communications and Mobile Computing}, vol.~9, no.~4, pp. 529--542, Apr. 2009.

\bibitem{imperfectcsi}
R.~Devarajan, A.~Punchihewa, and V.~K. Bhargava, ``Energy-aware power
  allocation in cooperative communication systems with imperfect {CSI},''
  \emph{IEEE Trans. Commun.}, vol.~61, no.~6, pp. 1633--1639, May 2013.

\bibitem{simeone}
O.~Simeone, Y.~Bar-Ness, and U.~Spagnolini, ``Stable throughput of cognitive
  radios with and without relaying capability,'' \emph{IEEE Trans. Commun.},
  vol.~55, no.~12, pp. 2351--2360, Dec. 2007.

\bibitem{khattab}
M.~Elsaadany, M.~Abdallah, T.~Khattab, M.~Khairy, and M.~Hasna, ``Cognitive
  relaying in wireless sensor networks: Performance analysis and
  optimization,'' in \emph{Proc. IEEE GLOBECOM}, Dec. 2010, pp. 1--6.

\bibitem{erph}
S.~Kompella, G.~Nguyen, J.~Wieselthier, and A.~Ephremides, ``Stable throughput
  tradeoffs in cognitive shared channels with cooperative relaying,'' in
  \emph{Proc. IEEE INFOCOM}, Apr. 2011, pp. 1961--1969.

\bibitem{su2011active}
W.~Su, J.~D. Matyjas, and S.~Batalama, ``Active cooperation between primary
  users and cognitive radio users in heterogeneous ad-hoc networks,''
  \emph{IEEE Trans. Signal Process.}, vol.~60, no.~4, pp. 1796--1805, Apr.
  2012.

\bibitem{rong2012cooperative}
B.~Rong and A.~Ephremides, ``Cooperative access in wireless networks: stable
  throughput and delay,'' \emph{IEEE Trans. Inf. Theory}, vol.~58, no.~9, pp.
  5890--5907, Sept. 2012.

\bibitem{chang2011cooperative}
H.-B. Chang and K.-C. Chen, ``Cooperative spectrum sharing economy for
  heterogeneous wireless networks,'' in \emph{IEEE GLOBECOM Workshops (GC
  Wkshps)}, 2011, pp. 458--463.

\bibitem{6504211}
H.~Roh, C.~Jung, W.~Lee, and D.-Z. Du, ``A stackelberg game for cooperative
  cognitive radio network with active sus,'' in \emph{ICNC}, Jan. 2013.

\bibitem{krikidis2009protocol}
I.~Krikidis, J.~Laneman, J.~Thompson, and S.~McLaughlin, ``Protocol design and
  throughput analysis for multi-user cognitive cooperative systems,''
  \emph{IEEE Trans. Wireless Commun.}, vol.~8, no.~9, pp. 4740--4751, Sept.
  2009.

\bibitem{sadek}
A.~Sadek, K.~Liu, and A.~Ephremides, ``Cognitive multiple access via
  cooperation: protocol design and performance analysis,'' \emph{IEEE Trans.
  Inf. Theory}, vol.~53, no.~10, pp. 3677--3696, Oct. 2007.

\bibitem{loynes1962stability}
R.~Loynes, ``The stability of a queue with non-independent inter-arrival and
  service times,'' in \emph{Proc. Cambridge Philos. Soc}, vol.~58, no.~3.\hskip
  1em plus 0.5em minus 0.4em\relax Cambridge University Press, 1962, pp.
  497--520.

\bibitem{riihonen2011mitigation}
T.~Riihonen, S.~Werner, and R.~Wichman, ``Mitigation of loopback
  self-interference in full-duplex mimo relays,'' \emph{IEEE Trans. Signal
  Process.}, vol.~59, no.~12, pp. 5983--5993, Mar. 2011.

\bibitem{bletsas2006simple}
A.~Bletsas, A.~Khisti, D.~P. Reed, and A.~Lippman, ``A simple cooperative
  diversity method based on network path selection,'' \emph{IEEE J. Sel. Areas
  Commun.}, vol.~24, no.~3, pp. 659--672, Aug. 2006.

\bibitem{scheuer1988reliability}
E.~M. Scheuer, ``Reliability of an m out of n system when component failure
  induces higher failure rates in survivors,'' \emph{IEEE Trans. Rel.},
  vol.~37, no.~1, pp. 73--74, 1988.

\end{thebibliography}
\end{document}